\documentstyle[12pt]{article}
\global\parskip 6pt

\normalsize

\def\br{\begin{thebibliography}{abc}}
\def\er{\end{thebibliography}}

\def\be{\begin{equation}}
\def\ee{\end{equation}}
\def\bea{\begin{eqnarray}}
\def\eea{\end{eqnarray}}

\begin{document}
\begin{flushright}
\hfill{SINP-TNP/06-26}\\
\end{flushright}
\vspace*{1cm}
\thispagestyle{empty}
\centerline{\large\bf Black Holes, Holography and Moduli Space Metric}
\bigskip
\begin{center}
Kumar S. Gupta\footnote{Email: kumars.gupta@saha.ac.in}\\
\vspace*{.2cm}
{\em Theory Division,\\ Saha Institute of Nuclear Physics\\
1/AF Bidhannagar, Calcutta - 700064, India}\\
\vspace*{.5cm}
Siddhartha Sen\footnote{Email: sen@maths.ucd.ie}\\
\vspace*{.2cm}
{\em School of Mathematical Sciences}\\
{\em UCD, Belfield, Dublin 4, Ireland}\\ 
\vspace*{.1cm}
{\em and}\\
\vspace*{.1cm}
{\em Department of Theoretical Physics}\\
{\em Indian Association for the Cultivation of Science}\\
{\em Calcutta - 700032, India}\\
\end{center}
\vskip.5cm

\begin{abstract}

String theory can accommodate black holes with the black hole parameters
related to string moduli. It is a well known but remarkable feature that the
near horizon geometry of a large class of black holes arising from string
theory contains a BTZ part.  A mathematical theorem (Sullivan's Theorem)
relates the three dimensional geometry of the BTZ metric to the conformal
structures of a two dimensional space, thus providing a precise kinematic
statement of holography. Using this theorem it is possible to argue that the
string moduli space in this region has to have negative curvature from the
BTZ part of the associated spacetime. This is consistent with a recent
conjecture of Ooguri and Vafa on string moduli space.

\end{abstract}
\vspace*{.3cm}
\begin{center}
October 2006
\end{center}
\vspace*{1.0cm}


\newpage

\baselineskip 18pt

String theory suggests that a consistent quantum theory of gravity should admit a holographic description. While a full understanding of quantum gravity is yet to emerge, black hole physics provides a glimpse into various quantum aspects of gravity. It is thus natural to ask what, if any, conclusions can be drawn regarding the metric of string moduli space from the holographic property of black holes. This is the issue we address in this Letter. We would argue that in certain regions of the moduli space of string theories, which corresponds to a large class of black holes, there exists a precise kinematical statement of holography. Our strategy relies on the observation that in certain parts of the string moduli space, the near horizon geometry contains a part with the causal structure of a BTZ black hole \cite{banados}. This observation has been very successful in the microscopic derivation of the Bekenstein-Hawking entropy for a large class of black holes \cite{maldastrom,physrept,peet}, including that for four and five dimensional non-supersymmetric backgrounds \cite{cvetic,skenderis}, e.g. Schwarzschild black hole. For all these backgrounds, which includes black holes of astrophysical interests, there exists a mathematical theorem (Sullivan's theorem) \cite{sull} that provides a precise kinematical statement of holography. Using this theorem we can argue that the metric of the string moduli space corresponding to the BTZ part of the spacetime has to have negative curvature. This precise result is consistent with a recent conjecture of Ooguri and Vafa about the curvature of the string moduli space \cite{vafa}.

We start by briefly recalling the string theories where the near horizon region has the causal structure of a BTZ black hole. The classic example is provided by the ten dimensional type IIB string theory compactified on $AdS_3 \times S^3 \times M^4$ \cite{maldastrom}. Using the notation of \cite{physrept}, the near horizon geometry of the solutions associated to $Q_1$ $D1$-branes and $Q_5$ $D5$-branes in the string frame is given by 
\be
\frac{ds^2}{\alpha^{\prime}} = \frac{U^2}{g_6 \sqrt{Q_1 Q_5}} \left (-dt^2 + (dx_5)^2 \right ) + 
g_6 \sqrt{Q_1 Q_5} \frac{dU^2}{U^2} + g_6 \sqrt{Q_1 Q_5} d \Omega_3^2
\ee
where $g_6$ is the six dimensional string coupling, $(2 \pi \alpha^{\prime})^{-1}$ is the string tensions and $l_s$ is the string length with
$l_s^2 = \alpha^{\prime}$. In the near horizon limit $U = \frac{r}{\alpha^{\prime}}$ is kept fixed as $\alpha^{\prime} \rightarrow 0$. The string moduli are fixed by the charges $Q_1$ and $Q_5$. 
The full geometry contains a contribution from the $M^4$ factor which has not been written above. The metric in (1) represents $AdS_3 \times S^3$ with $SL(2,R) \times SL(2,R)$ as isometry group. 
The BTZ metric is obtained from $ADS_3$ with a global identification by a fixed element of the isometry group \cite{banados,maldastrom} and the BTZ black hole parameters are identified in terms of the moduli. The above solution is supersymmetric in nature.

Another class of examples involve considering near extremal four dimensional rotating black hole solutions of toroidally compactified string theory, which admits five dimensional embeddings as rotating black strings \cite{cvetic}. The near extremal Kerr-Newman black hole falls in this category. The near-horizon geometry in this case is $AdS_3 \times S^2$, and the BTZ black hole is obtained with the usual global identifications of $Ads_3$.

The third class of examples include non-extremal and non-supersymmetric black holes in four and five dimensions which admit embedding into M-theory, and a series of U-dualities are then used to map these into configurations containing a BTZ black hole \cite{skenderis}. This includes the interesting example of four dimensional Schwarzschild black hole. The point of view here is that the U-duality transformations are analogous to gauge transformations which do not change the physical content of the system. By using the U-duality, the four and five dimensional black holes considered here are analyzed in the ``BTZ gauge".

Finally, it may be noted that in certain regions of the moduli space, the BTZ black hole arises directly as a solution of string theory \cite{horo}. Thus, by restricting attention to certain specific regions of the string moduli space and by use of dualities, the near-horizon geometry of a large class of black holes, including the ones of astrophysical interest, can be shown to contain a BTZ part. All the above examples were discussed in the literature mainly in the context of the calculation of black hole entropy. In fact, most of the effective string models used to calculate black hole entropy relied on having a BTZ geometry present in the system \cite{peet}. In all these cases, the BTZ black hole parameters, including the horizon radii, are determined from the string moduli.

As a next step we recall the key points which relates the BTZ black hole to a precise kinematical notion of holography. The basic feature of the BTZ
black hole which allows this to happen is that the Euclidean BTZ is 
a locally isomorphic to the hyperbolic 3-manifold $H^3$ which is geometrically 
finite \cite{sen}. The three
dimensional hyperbolic structures for such a manifold, according to
Sullivan's theorem \cite{sull}, are in 1-1 correspondence with the two
dimensional conformal structures of its boundary.  More precisely, if
$K$ is a geometrically finite hyperbolic  3-manifold with boundary then
 Sullivan's theorem states that as 
long as $K$ admits one hyperbolic realization, there is a 1-1 
correspondence between hyperbolic structures on $K$ and conformal 
structures on its boundary $\partial K$, the latter being the 
Teichmuller space of the boundary $\partial K$. This is nothing but a
precise mathematical statement of holography for the BTZ black hole
\cite{sen,man}. It is thus evident that for the very large class of black holes discussed above, which contain a BTZ geometry in the near-horizon region, the Sullivan's theorem would ensure this precise kinematical notion of holography would hold. The notion of holography here is kinematical as no detailed dynamical information about the theory is required to establish this correspondence. This is reminiscent of the "kinematical" nature of the entropy \cite{skenderis} calculated for the same class of black holes where the knowledge of the underlying CFT is enough to produce the Bekenstein-Hawking formula, without precise knowledge of the dynamics of the associated degrees of freedom \cite{ent}.

In order to proceed, note that the boundary of the BTZ black hole has the topology of
$T^2$ and the corresponding Teichmuller space is given by a the fundamental 
region of the complex variable $\tau$. In other words, two Teichmuller 
parameters $\tau$ and $\tau^{\prime}$ are equivalent if 
\be
\tau^{\prime} = \frac{a \tau + b}{c \tau + d}, ~~ ad - bc = 1
\ee
and $a,b,c,d \in Z$. The transformation in (2) is nothing but the action 
of the modular group on $\tau$ generated by the operations
\bea
S &:& \tau \rightarrow - \frac{1}{\tau} \nonumber \\
T &:& \tau \rightarrow \tau + 1.
\eea
In terms of the horizon radii $r_+$ and $r_-$ of the Euclidean BTZ black hole, 
the effective action of the modular group can be written as 
\bea
S &:& r_+ \leftrightarrow r_- \nonumber \\
T &:& r_+ \rightarrow r_+, ~~ r_- \rightarrow r_- +  r_+.
\eea 
This establishes the fact the that Riemann surface describing the boundary of the Euclidean BTZ has analytic properties. 

As mentioned before, the BTZ black hole was appears as a part of the near horizon geometry only in specific regions of the string moduli space. Indeed, the radii $r_+$ and $r_-$ of the BTZ black hole, as well as the corresponding Teichmuller parameter $\tau$, are fixed in terms of the string moduli. The tori obtained under the action of $S$ and $T$ in (4) are equivalent as Riemann surfaces. This implies that the metric in the region(s) of the moduli space where the BTZ black hole appears must also be invariant under $SL(2,Z)$. The moduli space metric with such an $SL(2,Z)$ invariance can be obtained by first constructing an $SL(2,R)$ invariant metric on which the $SL(2,Z)$ invariance can be imposed. To do this first note that under the SL(2,R) transformation 
\be
\tau \rightarrow \frac{\alpha \tau + \beta}{\gamma \tau + \delta},
\ee
where $\alpha \delta - \beta \gamma =1$ and  $ \alpha, \beta, \gamma, \delta \in R$, we have 
\be
d \tau \rightarrow \frac{d \tau}{{(\gamma \tau + \delta)}^2}.
\ee
It is then easy to check that metric
\be
ds^2 = \frac{d \tau  d {\bar{\tau}}}{{{\mathrm Im} \tau}^2}
\ee
is $SL(2,R)$ invariant. This can be interpreted as the metric in the region of the string moduli space for which the corresponding near horizon geometry contains a BTZ part. It is interesting to note that this metric has negative scalar curvature \cite{borel}.

In summary, based on the observation that in certain regions of the string moduli space the near-horizon geometry contains a BTZ part, and using a precise notion of holography for BTZ black holes following from Sullivan's theorem, we have argued that the corresponding regions of the string moduli space must admit a metric with negative curvature. This observation is consistent with one of the conjectures on the geometry of the moduli parametrizing the string landscape \cite{vafa}. Moreover, our observations provide a kinematical notion of holography for a large class of black holes, including the non-extremal and non-supersymmetric backgrounds, which are of phenomenological interest.


\bibliographystyle{unsrt}

\end{document}